\documentclass[preprint,preprintnumbers,amsmath,amssymb]{revtex4}
\usepackage{graphicx}
\begin{document}

\title{The continuum limit of the Bell model}

\author{Samuel Colin}
 \email{colin@fyma.ucl.ac.be}
\affiliation{FYMA, University of Louvain-la-Neuve, Belgium}

\date{\today}% It is always \today, today,
             %  but any date may be explicitly specified

\begin{abstract}
In a paper entitled \textit{Beables for Quantum Field Theory},
John Bell has shown that it was possible to build a realistic
interpretation of any hamiltonian lattice quantum field theory
involving Fermi fields. His model was constructed along the ideas
he used to present the de Broglie-Bohm pilot wave theory. However,
the beable (or element of reality) is now the fermion number
density, which is not a particle density, as in the de
Broglie-Bohm pilot wave theory. The model is stochastic but Bell
thought that it would become deterministic in the continuum limit.
We show that it is indeed the case, under an assumption about the
physical state of the universe, which follows naturally from the
Bell model. Moreover, the continuum model can be established
directly. The assumption is that the universe is in a state
obtained from the positronic sea (all positron states occupied) by
creating a finite number of negative charges. The physical
interpretation is the following: the negative charges are in
motion in the positronic sea and their positions are the beables
of the Bell model. The velocity laws we obtain for the motion of
the negative charges are very similar to those given by Bohm and
his co-workers for free relativistic fermions (first
quantization). The Bell model is non-local (it is unavoidable); we
show it explicitly in the simplest case. Under the previous
assumption about the state of the universe, and for quantum field
theories involving only Fermi fields, wave functions can be
defined, and calculations can be performed as in non-relativistic
quantum mechanics, since we stay in a sector of the Fock space
with a fixed fermion number.
\end{abstract}

\maketitle

\section{INTRODUCTION}
At the time Bell wrote his paper \cite{bell1}, Bohm had already
shown that it was possible to build a realistic interpretation of
any bosonic quantum field theory \cite{bohm1}. To achieve that
goal, Bohm took the field as the beable (or element of reality),
however he was not able to do the same for fermions. The aim of
Bell was then to show that it was also possible to build a
realistic interpretation of any fermionic quantum field theory,
along the pilot-wave ideas given by de Broglie and later by Bohm.
Bell managed doing so but he took a really different beable: the
fermion number density. It is quite different from the
non-relativistic pilot-wave theory, whose beables are the
positions of the particles. The model is also formulated on a
spatial lattice (space is discrete but time remains continuous).
His model is stochastic, but he suspected that the theory would
become deterministic in the continuum limit. The crucial point is
that the fermion number density is not a particle density. It is
in fact related to the charge density.

To deal with fermions on a spatial lattice, we have used the
Banks-Susskind-Kogut theory of staggered fermions \cite{kbs}. It
is the best way to eliminate the fermion doubling problem. Taking
the continuum limit of the Bell model for a staggered lattice, we
show that the theory becomes indeed deterministic in that limit.
The velocity laws we obtain are very similar to those given by
Bohm and co-workers for free relativistic fermions (first
quantization). We have worked with a one-dimensional lattice to
simplify the expressions but we think that it could be extended to
the three-dimensional case without difficulty. This remark is
justified since we show that the continuum model can also be
established directly.

Finally, we will study in more details the fact that any purely
fermionic quantum field theory can be brought to a particular
sector of the Fock space, with a fixed fermion number, allowing us
to define wave functions and to perform calculations in the same
way as they are done in non-relativistic quantum mechanics. We
will also study entanglement and non-locality in the simplest case
(two quanta of the fermion number).
\section{\label{sec:level1}THE BELL MODEL}

\subsection{\label{sec:level2}Ontology}
Three-space continuum is replaced by a finite lattice, whose sites
are labelled by an index
\begin{equation*}
l=1,2,\ldots,L~.
\end{equation*}
The fermion number density is the operator
\begin{equation*}
F(l)=\psi^\dagger(l)\psi(l)=\psi_1^\dagger(l)\psi_1(l)+\ldots+\psi^\dagger_4(l)\psi_4(l)~.
\end{equation*}
Since
\begin{equation}\label{comm}
[F(k),F(l)]=0~~\forall~k,l~\in~\{1,2,\ldots,L\}~,
\end{equation}
it is possible to define eigenstates of the fermion number
density:
\begin{equation*}
F(l)|n,q\rangle=f(l)|n,q\rangle~,
\end{equation*}
where $q$ are eigenvalues of observables $Q$ such that
$\{F(1),F(2),\ldots,F(L),Q\}$ is a complete set of observables,
and $n$ is a fermion number density configuration
($n=\{f(1),f(2),\ldots,f(L)\}$). Eigenvalues $f(l)$ belong to
$\{0,1,2,3,4\}$.

Thus we can imagine that the universe is in a definite fermion
number density configuration $n(t)$ at each time $t$, and that a
measurement of the fermion number density at time $t$ would simply
reveal the configuration $n(t)$. In Bell's words, the fermion
number density is given the beable status.

The second element, in the description of the universe, is the
pilot-state $|\Psi(t)\rangle$. Hence the universe, at time $t$, is
completely described by the couple $(|\Psi(t)\rangle,n(t))$.
\subsection{\label{sec:motion}Equations of motion}
For the pilot-state, the Schr{\"o}dinger equation is retained:
\begin{equation*}
i\frac{d|\Psi(t)\rangle}{dt}=H|\Psi(t)\rangle~.
\end{equation*}
An equation of motion for $n(t)$ must be added (it is called the
velocity-law in pilot-wave theories). Call $P_m(t)$ the
probability for the universe to be in configuration $m$ at time
$t$. Then the velocity-law must be such that the relation
\begin{equation}\label{equi}
P_m(t)=\sum_{q}|\langle m,q|\Psi(t)\rangle|^2
\end{equation}
holds for any time $t$, in order to reproduce the predictions of
orthodox quantum field theory. Since the configuration space is
discrete, it is impossible to find a deterministic velocity-law.
Instead jump-rates have to be defined. Call $T_{nm}(t)$ the
jump-rate for the transition $m\rightarrow n$ at time $t$, for
$m\neq n$ ($T_{mm}(t)=0~\forall m$). In other words, $T_{nm}(t)$
is the probability density (probability by unit of time) for the
universe to jump in configuration $n$, knowing that the universe
is in configuration $m$ at time $t$. The probability for the
universe to stay in configuration $m$ at time $t+dt$ is obtained
by the normalization condition and is equal to
\begin{equation}
1-\sum_{n}T_{nm}(t)dt~,
\end{equation}
for $dt$ small enough. Eq.~(\ref{equi}) is assumed to be true for
an initial time $t_0$; then the constraint on the stochastic
velocity-law becomes
\begin{equation}\label{equi2}
\frac{dP_m(t)}{dt}=\sum_{q}\frac{d}{dt}|\langle
m,q|\Psi(t)\rangle|^2~.
\end{equation}
Let's calculate the first member:
\begin{equation}
P_m(t+dt)=\sum_{n}T_{mn}(t)P_n(t)dt+(1-\sum_{n}T_{nm}(t))P_m(t)dt~,
\end{equation}
from which follows
\begin{equation}
\frac{dP_m(t)}{dt}=\lim_{dt\rightarrow
0}\frac{P_m(t+dt)-P_m(t)}{dt}=\sum_{n}(T_{mn}(t)P_n(t)-T_{nm}(t)P_m(t))~.
\end{equation}
Now let's calculate the second member of equation (\ref{equi2}).
With the help of the Schr\"odinger equation, we have
\begin{align}
\frac{d|\langle m,q|\Psi(t)\rangle|^2}{dt}=&\langle
m,q|\Psi(t)\rangle\langle\Psi(t)|iH|m,q\rangle+
\langle m,q|-iH|\Psi(t)\rangle\langle\Psi(t)|m,q\rangle&\\
=&2\mathfrak{Re}[\langle m,q|-iH|\Psi(t)\rangle\langle\Psi(t)|m,q\rangle]&\\
=&2\sum_{n,p}\mathfrak{Re}[\langle\Psi(t)|m,q\rangle\langle
m,q|-iH|n,p\rangle\langle n,p|\Psi(t)\rangle]~.&
\end{align}
Hence the constraint on the velocity law is
\begin{equation}
\sum_{n}(T_{mn}(t)P_n(t)-T_{nm}(t)P_m(t))
=2\sum_{n,p,q}\mathfrak{Re}[\langle\Psi(t)|m,q\rangle\langle
m,q|-iH|n,p\rangle\langle n,p|\Psi(t)\rangle]~,
\end{equation}
or
\begin{equation}
T_{mn}(t)P_n(t)-T_{nm}(t)P_m(t)=2\sum_{p,q}\mathfrak{Re}[\langle\Psi(t)|m,q\rangle\langle
m,q|-iH|n,p\rangle\langle n,p|\Psi(t)\rangle]~.
\end{equation}
If one takes for the following definition for the jump-rates:
\begin{align}
&T_{mn}(t)=\frac{J_{mn}(t)}{P_n(t)}& &\text{if}~J_{mn}\geq 0~,&\\
&T_{mn}(t)=0& &\text{otherwise}~,&
\end{align}
where
\begin{equation}\label{tc}
J_{mn}(t)=2\sum_{p,q}\mathfrak{Re}[\langle\Psi(t)|m,q\rangle\langle
m,q|-iH|n,p\rangle\langle n,p|\Psi(t)\rangle]~,
\end{equation}
then Eq.~(\ref{equi2}) is satisfied.
\subsection{\label{sec:comments}Comments}
We would like to make some remarks about the fermion number
density. First, eigenstates of the fermion number density are also
eigenstates of the charge density, which is
\begin{equation*}
-e:\psi^\dagger(l)\psi(l):~,
\end{equation*}
at least if we consider only electrons and positrons. The fermion
number
\begin{equation*}
F=\sum_{s=1}^{s=2}\int
d^3\vec{p}~[c^\dagger_s(\vec{p})c_s(\vec{p})+d_s(\vec{p})d^\dagger_s(\vec{p})]~,
\end{equation*}
is not the particle number, which is
\begin{equation*}
N=\sum_{s=1}^{s=2}\int
d^3\vec{p}~[c^\dagger_s(\vec{p})c_s(\vec{p})+d^\dagger_s(\vec{p})d_s(\vec{p})]~.
\end{equation*}
In fact the fermion number density does not commute with the
particle number; it is possible to find well-behaved functions
$f(\vec{x})$ such that
\begin{equation*}
[\int d^3\vec{x}
f(\vec{x})\psi^\dagger(\vec{x})\psi(\vec{x}),N]\neq 0~.
\end{equation*}
The proof is given in appendix \ref{sec:appa}.

Thus the charge density does not commute with the particle number
either. A measurement of the charge contained in any finite region
including the coordinate $\vec{x}_0$, with value$-e$, is never an
electron. In fact, as we will see, it is a superposition
containing one electron, two electrons and one positron, three
electrons and two positrons, and so on. On one hand, it is
disturbing, since the tracks observed in bubble chambers are said
to represent electrons or positrons. On the other hand, when the
electromagnetic field is taken into account, it is quite natural,
since measurement of localized properties involve high energy
radiation, and thus that can lead to pair creations. But these are
are just few remarks to draw attention to the interpretation
problems one has to cope with.
\section{The Dirac theory in a $1+1$ space-time}
\subsection{The Dirac equation}
\subsubsection{Solutions of the Dirac equation}
The hamiltonian is
\begin{equation}
H=\alpha p+\beta m~.
\end{equation}
Since $H^\dagger=H$ and $H^2=p^2+m^2$, we obtain the following relations:
\begin{align}
&\alpha^\dagger=\alpha& &\beta^\dagger=\beta& &\{\alpha,\beta\}=0& &\alpha^2=\beta^2=1&~.
\end{align}
The smallest dimension for a representation of that algebra is two. For example:
\begin{align}
&\beta=\begin{pmatrix}
  1 & 0 \\
  0 & -1 \\
\end{pmatrix}&
&\alpha=\begin{pmatrix}
  0 & 1 \\
  1 & 0 \\
\end{pmatrix}&~.
\end{align}
Thus we have a spinor of dimension $2$, $\psi(t,x)$, which is solution of
\begin{equation}
i\frac{\partial\psi(t,x)}{\partial t}=-i\alpha\frac{\partial\psi(t,x)}{\partial x}+m\beta\psi(t,x)~.
\end{equation}
This equation can be rewritten in a covariant form, by introducing the $\gamma$ matrices, defined by
\begin{align}
&\gamma^0=\beta=\gamma_0& &\gamma^1=\alpha=-\gamma_1&~.
\end{align}
That implies
\begin{equation}
(i\gamma^\mu\partial_\mu-m)\psi(t,x)=0~.
\end{equation}
One can verify that the $\gamma$ matrices satisfy the following relations:
\begin{align}
&\{\gamma^\mu,\gamma^\nu\}=2g^{\mu \nu}&
&{\gamma^\mu}^\dagger=\gamma^0\gamma^\mu\gamma^0&~.
\end{align}
Each component of the spinor is a solution of the Klein-Gordon equation:
\begin{equation}
(\square+m^2)\psi(t,x)=0~.
\end{equation}
Now we search for free solutions; the most general form is thus
\begin{align}
&u(p)e^{-i{E_p}t}e^{ipx}& &v(p)e^{i{E_p}t}e^{-ipx}&~,
\end{align}
with $p~\in~\mathfrak{R}$ and $E_p=\sqrt{p^2+m^2}$. We find that
\begin{align}
&\psi^p_+(t,x)=u(p)e^{-iE_pt}e^{ipx}=\begin{pmatrix}
  1\\
  \frac{p}{m+E_p} \\
\end{pmatrix}e^{-iE_p t}e^{ipx}&\\
&\psi^p_-(t,x)=u(p)e^{iE_p t}e^{-ipx}=\begin{pmatrix}
  \frac{p}{m+E_p} \\ 1\\
\end{pmatrix}e^{iE_p t}e^{-ipx}&
\end{align}
are solutions of Dirac equation, respectively of positive and negative energy.
To obtain an interpretation of the theory, we need a conserved current, whose temporal component is
positive:
\begin{align}
&\partial_\mu j^\mu=0& &\text{with } j^0\geq 0&~.
\end{align}
The current
\begin{equation}
j^\mu=\bar{\psi}\gamma^\mu\psi~,
\end{equation}
with $\bar{\psi}=\psi^\dagger\gamma^0$, is suitable. Spinors normalization still remains to be discussed. Assume that the universe
is a box of volume $V$, in a inertial frame $\Sigma$ where the momentum of the free particle is $p$. Then the quantity
\begin{equation*}
\int_{V}dx {j^0}(t,x)
\end{equation*}
must be equal to $1$ in every inertial frame. That means that the
spinors $u(p)$ and $v(p)$ must be normalized to
$\frac{1}{V_0}\frac{E_p}{m}$, where $V_0$ is the volume of the
universe in an inertial frame where the particle is at rest.
\subsubsection{Physical interpretation}
Dirac obtained a conserved current, whose time-component is
positive, a task that was impossible with the Klein-Gordon
equation. But the negative energy states are still there. Once
interactions are taken into account, that would lead to the
instability of the hydrogen atom, for example. To avoid this,
Dirac assumed that all negative energy states were occupied. Hence
a positive energy electron cannot transit to a negative energy
state, due to Pauli exclusion principle. That state of lowest
energy is called the Dirac sea. It is impossible to distinguish it
from a state where no electrons are present. The absence of a
negative energy state of momentum $p$ ( a hole in the Dirac sea)
would be seen as a particle of positive energy $\sqrt{p^2+m^2}$,
momentum $-p$ and charge $e$. That led to the prediction of
anti-particles known as positrons.
\subsection{The Dirac quantum field theory} The
first step, in the construction of the corresponding quantum field
theory, is to obtain a classical relativistic action, from which
we can obtain the Dirac equation, by using the least action
principle. The following action is suitable:
\begin{equation}
S[\psi,\psi^\dagger]=\int
dxdt\bar{\psi}(t,x)(i\gamma^\mu\partial_\mu-m)\psi(t,x)~.
\end{equation}
It is not hermitian but it can be rewritten as
\begin{equation*}
S=S_h+\int\partial_\mu J^\mu~,
\end{equation*}
where $S_h$ is hermitian, and the last term can be dropped. We get
the momenta conjugate to the fields:
\begin{align}
&\pi_a(t,x)=\frac{\partial\mathcal{L}(t,x)}{\partial(\dot{\psi}_a(t,x))}=i\psi^*_a(t,x)&
&\pi^*_a(t,x)=\frac{\partial\mathcal{L}(t,x)}{\partial(\dot{\psi}^*_a(t,x))}=0&~.
\end{align}
The next step is quantization, according to the canonical
equal-time anti-commutation relations: classical fields become
quantum fields, obeying the relation
\begin{align*}
&\{\psi_a(t,x),\pi_b(t,y)\}=i\delta(x-y)\delta_{a b}&
&\{\psi_a(t,x),\psi_b(t,y)\}=0& &\{\pi_a(t,x),\pi_b(t,y)\}=0~.&
\end{align*}
Those relations can be rewritten as
\begin{align}\label{etacr}
&\{\psi_a(t,x),\psi^\dagger_b(t,y)\}=\delta(x-y)\delta_{a b}&
&\{\psi_a(t,x),\psi_b(t,y)\}=0~.&
\end{align}
Since the quantum field $\psi(t,x)$ satisfies the Dirac equation,
it is a superposition of free solutions with operators as
coefficients:
\begin{equation}\label{spi1}
\psi(t,x)=\frac{1}{\sqrt{2\pi}}\int
{dp}[c(p)u(p)e^{-i{E_p}t}e^{ipx}+\zeta(-p)v(p)e^{i{E_p}t}e^{-ipx}]~,
\end{equation}
\begin{equation}\label{spi2}
\psi^\dagger(t,x)=\frac{1}{\sqrt{2\pi}}\int
{dp}[c^\dagger(p)u^T(p)e^{i{E_p}t}e^{-ipx}+\zeta^\dagger(-p)v^T(p)e^{-i{E_p}t}e^{ipx}]~.
\end{equation}
$c$, $\zeta$, $c^\dagger$ and $\zeta^\dagger$ are operators
satisfying unknown anti-commutation relations, that must be chosen
in order to regain the equal-time anti-commutation relations (Eq.
(\ref{etacr})). Spinors are normalized to
\begin{align*}
&u^\dagger(p)u(p)=\frac{E_p}{m}&
&v^\dagger(p)v(p)=\frac{E_p}{m}&~.
\end{align*}
With the help of Eq. (\ref{spi1}) and Eq. (\ref{spi2}), we can
work out the equal-time anti-commutation relations. We have
\begin{align*}
\{\psi_a(t,x),\psi_b(t,y)\}=\frac{1}{2\pi}\int{dp}{dq}
[&\{c(p),c(q)\}e^{-i{E_p}t}e^{-i{E_q}t}e^{ipx}e^{iqy}u_a(p)u_b(q)&\\
+&\{\zeta(-p),\zeta(-q)\}e^{i{E_p}t}e^{i{E_q}t}e^{-ipx}e^{-iqy}v_a(p)v_b(q)&\\
+&\{c(p),\zeta(-q)\}e^{-i{E_p}t}e^{i{E_q}t}e^{ipx}e^{-iqy}u_a(p)v_b(q)&\\
+&\{\zeta(-p),c(q)\}e^{i{E_p}t}e^{-i{E_q}t}e^{-ipx}e^{iqy}v_a(p)u_b(q)]~.&
\end{align*}
If we take
\begin{align}\label{comm1}
&\{c(p),c(q)\}=0& &\{c(p),\zeta(q)\}=0& &\{\zeta(p),\zeta(q)\}=0&
&\forall~p,q~,&
\end{align}
then we obtain $\{\psi_a(t,x),\psi_b(t,y)\}=0$. The relation
$\{\psi_a(t,x),\psi^\dagger_b(t,y)\}=\delta(x-y)\delta_{a b}$
remains to be considered. With the help of Eqs (\ref{spi1}) and
(\ref{spi2}), we have
\begin{align*}
\{\psi_a(t,x),\psi^\dagger_b(t,y)\}=\frac{1}{2\pi}\int{dp}{dq}
[&\{c(p),c^\dagger(q)\}e^{-i{E_p}t}e^{i{E_q}t}e^{ipx}e^{-iqy}u_a(p)u_b(q)&\\
+&\{\zeta(-p),\zeta^\dagger(-q)\}e^{i{E_p}t}e^{-i{E_q}t}e^{-ipx}e^{iqy}v_a(p)v_b(q)&\\
+&\{c(p),\zeta^\dagger(-q)\}e^{-i{E_p}t}e^{-i{E_q}t}e^{ipx}e^{iqy}u_a(p)v_b(q)&\\
+&\{\zeta(-p),c^\dagger(q)\}e^{i{E_p}t}e^{i{E_q}t}e^{-ipx}e^{-iqy}v_a(p)u_b(q)]~.&
\end{align*}
Taking
\begin{align*}
&\{c(p),c^\dagger(q)\}=\frac{m}{E_p}\delta(p-q)&
&\{\zeta(p),\zeta^\dagger(q)\}=\frac{m}{E_p}\delta(p-q)&
&\{c(p),\zeta^\dagger(q)\}=0& &\forall~p,q~,&
\end{align*}
the relations (\ref{etacr}) are regained. We can also choose
\begin{equation}
\psi(t,x)=\frac{1}{\sqrt{2\pi}}\int
{dp}\sqrt{\frac{m}{E_p}}[c(p)u(p)e^{-iEt}e^{ipx}+\zeta(-p)v(p)e^{iEt}e^{-ipx}]~,
\end{equation}
with the following anti-commutation relations:
\begin{align*}
&\{c(p),c(q)\}=0& &\{c(p),\zeta(q)\}=0& &\{\zeta(p),\zeta(q)\}=0&
\\
&\{c(p),c^\dagger(q)\}=\delta(p-q)&
&\{\zeta(p),\zeta^\dagger(q)\}=\delta(p-q)&
&\{c(p),\zeta^\dagger(q)\}=0& &\forall~p,q~.&
\end{align*}
That is the choice we adopt. Now the observables can be expressed
in the momentum space. For the hamiltonian, we have
\begin{equation*}
H=\int {dx}\psi^\dagger(x)[-i\alpha\nabla+m\beta]\psi(x)=\int
{dp}\sqrt{p^2+m^2}[c^\dagger(p)c(p)-\zeta^\dagger(-p)\zeta(-p)]~.
\end{equation*}
The momentum is
\begin{equation*}
P=\int {dx}\psi^\dagger(x)[-i\nabla]\psi(x)=\int
{dp}p[c^\dagger(p)c(p)-\zeta^\dagger(-p)\zeta(-p)]~.
\end{equation*}
And the fermion number is
\begin{equation*}
F=\int
{dx}\psi^\dagger(x)\psi(x)=\int{dp}[c^\dagger(p)c(p)+\zeta^\dagger(-p)\zeta(-p)]~.
\end{equation*}
We can define a vacuum as a state annihilated by any operator
$c(p)$ or $\zeta(p)$; we call that state $|0_1\rangle$:
\begin{align*}
&c(p)|0_1\rangle=0& &\zeta(p)|0_1\rangle=0& &\forall p~.&
\end{align*}
Now it is clear that $c^\dagger(p)$ creates an electron of energy
$\sqrt{p^2+m^2}$ and momentum $p$, whereas $\zeta^\dagger(p)$
creates an electron of energy $-\sqrt{p^2+m^2}$ and momentum $p$,
and that the Dirac sea is the state
$|DS\rangle=\displaystyle\prod_{p}\zeta^\dagger(p)|0_1\rangle$.
Usually, everything is rewritten by introducing positrons, by
making the substitutions
\begin{align*}
&\zeta^\dagger(p)\rightarrow d(-p)& &\zeta(p)\rightarrow
d^\dagger(-p)~,&
\end{align*}
where $d^\dagger(p)$ is the operator that creates a positron of
momentum $p$ and energy $\sqrt{p^2+m^2}$. And we have to define
another vacuum, $|0_2\rangle$:
\begin{align*}
&c(p)|0_2\rangle=0& &d(p)|0_2\rangle=0& &\forall p~.&
\end{align*}
We have the relations
\begin{align*}
&|0_2\rangle=\prod_{p}\zeta^\dagger(p)|0_1\rangle&
&|0_1\rangle=\prod_{p}d^\dagger(p)|0_2\rangle~.&
\end{align*}
\section{\label{sec:ferlat}FERMIONS ON A LATTICE}
One-space continuum is replaced by a lattice of spacing $\delta$,
having $L=2N$ sites. The momentum space is also a lattice, having $2N$ sites and a
spacing $\frac{\pi}{N\delta}$. To get the lattice action, one makes the
following substitutions
\begin{align*}
&\int dx\rightarrow \delta\sum_{l} &
&\psi(t,x)\rightarrow\psi(t,j)& &\frac{\partial\psi(t,x)}{\partial
x}\rightarrow\frac{\psi(t,j+1)-\psi(t,j-1)}{2\delta}&
\end{align*}
in the continuum action
\begin{equation*}
S[\psi,\psi^\dagger]=\int dx
dt\bar{\psi}(t,x)[i\gamma^\mu\partial_\mu-m]\psi(t,x)~.
\end{equation*}
Doing so, we obtain
\begin{equation*}
S_{lat}[\psi,\psi^\dagger]=\delta\sum_{j}\int dt
[i\psi^\dagger(j)\partial_t\psi(j)+i\psi^\dagger(j)\alpha\frac{\psi(t,j+1)-\psi(t,j-1)}
{2\delta}-m\psi^\dagger(j)\beta\psi(j)]~.
\end{equation*}
We can eliminate the factor $\delta$, by making the substitution
\begin{equation*}
\psi(t,j)\rightarrow\frac{\psi(t,j)}{\sqrt{\delta}}~,
\end{equation*}
so that the lattice Dirac action is
\begin{equation*}
S_{lat}[\psi,\psi^\dagger]=\sum_{j}\int dt
[i\psi^\dagger(j)\partial_t\psi(j)+i\psi^\dagger(j)\alpha\frac{\psi(t,j+1)-\psi(t,j-1)}{2\delta}-m\psi^\dagger(j)\beta\psi(j)]~.
\end{equation*}
The least action principle gives the lattice Dirac equation (we
use periodic conditions on the boundaries $\psi_{-N}=\psi_{N}$):
\begin{equation*}
i\frac{\partial\psi(t,j)}{\partial t}=-i\alpha\frac{\psi(t,j+1)-\psi(t,j-1)}{2\delta}+m\beta\psi(t,j)~,
\end{equation*}
whose free solutions are
\begin{align}
&\psi^p_+(t,j)=u(p_{lat})e^{-iE_{lat}(p)t}e^{ipj\delta}=\begin{pmatrix}
  1\\
  \frac{p_{lat}}{m+E_{lat}(p)} \\
\end{pmatrix}e^{-iE_{lat}(p)t}e^{ipj\delta}&\\
&\psi^p_-(t,j)=u(p_{lat})e^{iE_{lat}t}e^{-ipj\delta}=\begin{pmatrix}
  \frac{p_{lat}}{m+E_{lat}(p)} \\ 1\\
\end{pmatrix}e^{iE_{lat}(p)t}e^{-ipj\delta}&~,
\end{align}
where $p_{lat}=\frac{\sin(p\delta)}{\delta}$ and
$E_{lat}(p)=\sqrt{p^2_{lat}+m^2}$. Now we turn to the
quantization. The anti-commutation relations become
\begin{equation*}
\{\psi_a(t,j),{\psi^\dagger}_b(t,k)\}=\delta^a_b\delta^j_k~,
\end{equation*}
and all other anti-commutators vanishing. Since it satisfies the
lattice Dirac equation, the quantum field $\psi(t,j)$ is a
superposition of free solutions, with operators as coefficients:
\begin{equation*}
\psi(t,j)=\sum_{p}\omega(p)[c(p)u(p_{lat})e^{-iE_{lat}(p)t}e^{ipj\delta}+d^\dagger(p)v(p_{lat})e^{iE_{lat}(p)t}e^{-ipj\delta}]~.
\end{equation*}
Again, $\omega(p)$ and the anti-commutation relations satisfied by
the operators $c$, $d$, $c^\dagger$ and $d^\dagger$, are
determined by the canonical equal-time anti-commutation relations.
But is not difficult to see that the hamiltonian is
\begin{equation*}
H=\sum_{p}\sqrt{\frac{\sin^2(p\delta)}{\delta^2}+m^2}[c^\dagger(p)c(p)-d(p)d^\dagger(p)]
\end{equation*}
Thus there are four states of energy $m$. Generally, every
eigenstate containing $n$ particles is degenerate, with degeneracy
$2^n$. The same problem occur in the propagator; it has four poles
and thus propagates twice more particles. It is called the fermion
doubling problem and there are many theories to deal with it (the
Wilson theory, the Banks-Susskind-Kogut theory of staggered
fermions, to mention the main ones). In the continuum limit
(lattice spacing going to zero and finite momentum), we have
\begin{equation*}
\frac{\sin(p\delta)}{\delta}\rightarrow p~,
\end{equation*}
so that the fermion doubling problem disappears. But that is not a
reason to ignore it in our calculations. In the one-dimensional
case, it is easier to use the Banks-Susskind-Kogut theory of
staggered fermions to overcome it.
\subsection{The Banks-Susskind-Kogut theory of staggered fermions}
The idea is to start from the previous theory, with a lattice
spacing equal to $2\delta$, and to say that there are two
superposed lattices, one where upper components of $\psi(t,j)$
live, and another one where lower components live. By moving the
lower lattice to the right, with a translation of magnitude
$\delta$, we obtain a theory of a complex field $\phi(l)$, over a
lattice containing twice more sites.

Since the part of the article which we are interested in is quite
small, we will just quote it \cite{kbs}:
\begin{quote}
Consider a spatial lattice (continuous time) with a lattice spacing $a$. Label the lattice sites with an integer $n$. There
will be a one-component fermion field $\phi(n)$ at each site $n$. $\phi(n)$ satisfies the anti-commutation
relation
\begin{align}\label{kbs1}
&\{\phi^\dagger(n),\phi(m)\}=\delta_{nm}& &\{\phi(n),\phi(m)\}=0&
\end{align}
$\phi(n)$ is related to a properly normalized continuum field
$\chi$ having canonical anti-commutation relations by
\begin{equation}\label{kbs2}
\phi(n)=\sqrt{a}\chi(x)
\end{equation}
Consider the hamiltonian
\begin{equation}\label{kbs3}
H=\frac{i}{2a}\sum_{n}[\phi^\dagger(n)\phi(n+1)-\phi^\dagger(n+1)\phi(n)]
\end{equation}
We claim that with a proper identification of a two-component fermion fields Eqs (\ref{kbs1})-(\ref{kbs3}) generate
the massless Dirac equation in the continuum limit. First compute
\begin{equation}\label{kbs4}
i[H,\phi(n)]=\dot{\phi}(n)=\frac{\phi(n+1)-\phi(n-1)}{2a}
\end{equation}
Note that the time dependence of $\phi(n)$ at even (odd) sites is determined by the spatial difference of
$\phi(n\pm 1)$ at odd (even) sites. So, to ensure finite time dependence in $\phi(n)$ at even (odd) sites, we must require
that the spatial dependence in $\phi(n)$ at odd (even) sites be smooth. Thus we defined a two-component field
$\psi(n)$ as follows:
\begin{equation*}
\psi=\begin{pmatrix}\psi_e \\ \psi_o\end{pmatrix}
\end{equation*}
\begin{equation}
\psi_e(n)=\phi(n), ~\text{ n even}
\end{equation}
\begin{equation*}
\psi_o(n)=\phi(n), ~\text{ n odd}.
\end{equation*}
Then the components of $\psi(n)$ satisfy the equations
\begin{align}\label{kbs6}
&\dot{\psi}_o=\frac{\Delta \psi_e}{\Delta x},& &\dot{\psi}_e=\frac{\Delta \psi_o}{\Delta x}&
\end{align}
where $\Delta$ indicates the discrete difference in Eq. (\ref{kbs4}). Note that Eq. (\ref{kbs6}) becomes the massless
Dirac equation in the continuum limit,
\begin{equation}
\frac{\partial}{\partial t}\psi=\begin{pmatrix} 0 & 1\\1 & 0\end{pmatrix}\frac{\partial}{\partial x}\psi
\end{equation}
in a standard basis where
\begin{equation*}
\gamma_0=\begin{pmatrix} 1 & 0\\0 & -1\end{pmatrix}
\end{equation*}
\end{quote}
If we consider the case of massive fermions, we just add a term $\displaystyle m\sum_{n}(-1)^n\phi^\dagger(n)\phi(n)$ to the massless
hamiltonian:
\begin{equation*}
H=-\frac{i}{2\delta}\sum_{n}[\phi^\dagger(n)\phi(n+1)-\phi^\dagger(n+1)\phi(n)]+\sum_{n}m(-1)^n\phi^\dagger(n)\phi(n)~.
\end{equation*}
There is a minus sign in the kinetic term, compared to the
expression in the Banks-Susskind-Kogut article, but that is just a
matter of conventions. Again we take the number of lattice sites
to be $2N$. Thus the momentum space is a lattice containing $2N$
sites, with spacing $\frac{\pi}{N\delta}$. What about the fermion
doubling problem? The momentum lattice is still
$\displaystyle\frac{\sin(p\delta)}{\delta}$, but the momentum
space is divided into two sub-spaces:
\begin{itemize}
\item[$\cdot$]the sub-space $\mathcal{P}_1$, where electrons live.
\item[$\cdot$]the sub-space $\mathcal{P}_2$, where positrons live.
\end{itemize}
And it is impossible to find $p,q\in\mathcal{P}_1$ (resp.
$\mathcal{P}_2$) such that
$\displaystyle\frac{\sin(p\delta)}{\delta}=\frac{\sin(q\delta)}{\delta}$.
It is quite straightforward to show if we think in terms of
degrees of freedom. Thus there are $N$ sites in $\mathcal{P}_1$
and $N$ sites in $\mathcal{P}_2$.
\subsubsection{Eigenstates of the fermion number density}
The fermion number density is the following operator:
\begin{equation*}
\phi^\dagger(l)\phi(l)~.
\end{equation*}
The fermion number density is an operator with positive
eigenvalues. Summing over all sites, we obtain the fermion number:
\begin{equation*}
\sum_{l}\phi^\dagger(l)\phi(l)=\sum_{p\in\mathcal{P}_1}c^\dagger(p)c(p)+\sum_{p\in\mathcal{P}_2}d(p)d^\dagger(p)~.
\end{equation*}
The eigenvalues of the fermion number range from $0$ to $2N$. The
state with the lowest fermion number is thus the positronic sea
(all positron states occupied):
\begin{equation*}
|0_1\rangle=\prod_{p\in\mathcal{P}_2}d^\dagger(p)|0_2\rangle~.
\end{equation*}
Since the eigenvalues of the fermion number density are positive,
there can be only one eigenstate of the fermion number density
with fermion number equal to $0$; it is thus the positronic sea:
\begin{equation*}
\phi^\dagger(l)\phi(l)|0_1\rangle=0~.
\end{equation*}
From the anti-commutation relation
\begin{equation*}
\{\phi(l),\phi^\dagger(k)\}=\delta_{lk}~,
\end{equation*}
we find that $\phi^\dagger(k)$ creates a quantum of the fermion
number at site $k$, whereas $\phi(k)$ destroys it. The eigenvalues
of $\phi^\dagger(l)\phi(l)$ belong to $\{0,1\}$. The positronic
sea is annihilated by any annihilator $\phi(k)$ (it could be
called the vacuum of the fermion number density). Eigenstates of
the fermion number density are thus generated by applying creators
at different sites.
\section{\label{sec:climit}THE CONTINUUM LIMIT}
The fermion number commutes with the hamiltonian
\begin{equation*}
[H,\sum_{l}\phi^\dagger(l)\phi(l)]=0~.
\end{equation*}
and any physical state is an eigenstate of the fermion number, so
there is an integer $\omega$ such that
\begin{equation*}
\sum_{l}\phi^\dagger(l)\phi(l)|\Psi(t)\rangle=\omega|\Psi(t)\rangle~.
\end{equation*}
for any time $t$ (with $\omega~\in~\{0,1,\ldots,2N\}$).
\subsection{A localized quantum}
Let's first consider the case of a localized quantum
\begin{equation*}
\sum_{l}\phi^\dagger(l)\phi(l)|\Psi(t)\rangle=|\Psi(t)\rangle~.
\end{equation*}
Since we are restricted to the sector of the Fock space with the
fermion number equal to $1$, we use a more convenient notation for
the eigenstates of $\phi^\dagger(l)\phi(l)$
\begin{align*}
&|k\rangle=\phi^\dagger(k)|0_1\rangle&
&\phi^\dagger(l)\phi(l)|k\rangle=\delta_{lk}|k\rangle&~.
\end{align*}
The pilot-state can be decomposed in the basis $\{|k\rangle\}$:
\begin{equation*}
|\Psi(t)\rangle=\sum_{k=-N}^{k=N-1}\Psi(t,k)|k\rangle~.
\end{equation*}
Assume that the beable at the initial time is
$n_l(t_0)=\delta_{lk}$, corresponding to the state
$|k\rangle=\phi^\dagger(k)|0_1\rangle$. How does it evolve with
time? The hamiltonian matrix element which appears in the
transition current (Eq. \ref{tc}) is
\begin{eqnarray*}
\langle l|-iH|k\rangle=\langle
0_1|-i\phi(l)H\phi^\dagger(k)|0_1\rangle=\frac{\delta_{l(k+1)}}{2\delta}-\frac{\delta_{l(k-1)}}{2\delta}~.
\end{eqnarray*}
Thus they are only two transitions which are allowed:
\begin{align*}
&J_{(k+1)k}=\frac{1}{\delta}\mathfrak{Re}[\Psi^*(t,k+1)\Psi(t,k)]&\\
&J_{(k-1)k}=-\frac{1}{\delta}\mathfrak{Re}[\Psi^*(t,k-1)\Psi(t,k)]&~.
\end{align*}
The quantum can only jump to the first neighbor site, either to
the right or the left. Using the relation
\begin{equation*}
\Psi^*(t,k-1)=\Psi^*(t,k+1)-2\delta\nabla\Psi^*(t,k)
\end{equation*}
and keeping only leading terms, we get:
\begin{align*}
&J_{(k+1)k}=\frac{1}{\delta}\mathfrak{Re}[\Psi^*(t,k+1)\Psi(t,k)]&\\
&J_{(k-1)k}=-\frac{1}{\delta}\mathfrak{Re}[\Psi^*(t,k+1)\Psi(t,k)]~.&
\end{align*}
These two currents have opposite signs; then there is only one
transition. The theory is thus deterministic. Let's consider the
case $\mathfrak{Re}[\Psi^*(t,k+1)\Psi(t,k)]\geq 0$. Then the
quantum moves towards the right and its velocity is
\begin{equation}\label{vlat}
v=\frac{\delta T_{(k+1)k}dt}{dt}=\frac{\delta
J_{(k+1)k}}{|\Psi(t,k)|^2}=\frac{\mathfrak{Re}[\Psi^*(t,k+1)\Psi(t,k)]}
{|\Psi(t,k)|^2}
\end{equation}
Now what about the continuum limit? Remember that even sites
correspond to upper components of the spinor, whereas odd sites
correspond to lower components of the spinor. In the continuum
limit, the couple $(\phi^\dagger(n),\phi^\dagger(n+1))$, for $n$
even, tends to the spinor
$(\psi^\dagger_{1}(x),\psi^\dagger_{2}(x))$. There are two sites
that converge to a given $x$. If $k$ was even, we still have to
consider its companion $k+1$. Then let's assume that the beable is
$\delta_{l(k+1)}$ at time $t_0$. One can find that the currents
are
\begin{align*}
&J_{(k+2)k+1}=-\frac{1}{\delta}\mathfrak{Re}[\Psi^*(t,k)\Psi(t,k+1)]&\\
&J_{(k)k+1}=\frac{1}{\delta}\mathfrak{Re}[\Psi^*(t,k)\Psi(t,k+1)]~.&
\end{align*}
We still consider the case
$\mathfrak{Re}[\Psi^*(t,k+1)\Psi(t,k)]\geq 0$. Then only the
current $J_{(k+2)k+1}$ survives and we find that
\begin{equation*}
J_{(k+1)k}=J_{(k+2)k+1}=\frac{1}{\delta}\mathfrak{Re}[\Psi^*(t,k+1)\Psi(t,k)]~.
\end{equation*}
The continuum current is just the continuum version of the sum of
the two lattice currents (the one for $k$ plus the other one for
$k+1$):
\begin{align*}
J(x,t)=&2\mathfrak{Re}[\langle \Psi(t)|\psi^\dagger_2(t,x)|0_1\rangle\langle 0_1|\psi_1(t,x)|\Psi(t)\rangle]&\\
=&\langle \Psi(t)|\psi^\dagger(t,x)|0_1\rangle\alpha\langle
0_1|\psi(t,x)|\Psi(t)\rangle&~.
\end{align*}
\subsection{Two localized quanta}
Now let's consider the case of two localized quanta:
\begin{equation*}
\sum_{l}\phi^\dagger(l)\phi(l)|\Psi(t)\rangle=2|\Psi(t)\rangle~.
\end{equation*}
We use the same notations: an eigenstate of
$\phi^\dagger(l)\phi(l)$ with eigenvalue
$n_l=\delta_{l{k_1}}+\delta_{l{k_2}}$ is denoted
\begin{align*}
&|k_1,k_2\rangle=\phi^\dagger(k_1)\phi^\dagger(k_2)|0_1\rangle&~,
\end{align*}
with $k_1<k_2$ and $k_1,k_2~\in~\{-N,-N+1,\ldots,N-1\}$,
corresponding to the state
$\phi^\dagger(k_1)\phi^\dagger(k_2)|0_1\rangle$. The most general
beable at the initial time is thus
$n_l(t_0)=\delta_{l{k_1}}+\delta_{l{k_2}}$. The hamiltonian matrix
element which appears in the transition current is
\begin{align*}
\langle l_1,l_2|-i{H}|k_1,k_2\rangle=&-\sum_{l}\langle
0_1|\phi(l_2)\phi(l_1)\phi^\dagger(l)\frac{\phi(l+1)-\phi(l-1)}{2\delta}\phi^\dagger(k_1)\phi^\dagger(k_2)|0_1\rangle&~.
\end{align*}
Using the canonical anti-commutation relations and the fact that
the positronic sea is annihilated by any $\phi(n)$, and taking
into account the constraints $k_1<k_2$ and $l_1<l_2$, one finds
that
\begin{equation*}
\langle
l_1,l_2|-i{H}|k_1,k_2\rangle=\frac{\delta_{l_2 (k_2+1)}\delta_{l_1 k_1}+\delta_{l_1 (k_1+1)}\delta_{l_2 k_2}
-\delta_{l_2 (k_2-1)}\delta_{l_1 k_1}-\delta_{l_1 (k_1-1)}\delta_{l_2 k_2}}{2\delta}~,
\end{equation*}
only if $k_2\neq k_1+1$, otherwise
\begin{equation*}
\langle
l_1,l_2|-i{H}|k_1,k_2\rangle=\frac{\delta_{l_2 (k_2+1)}\delta_{l_1 k_1}-\delta_{l_1 (k_1-1)}\delta_{l_2 k_2}}{2\delta}~.
\end{equation*}
The pilot-state can be decomposed in the basis
$\{\phi^\dagger(n_1)\phi^\dagger(n_2)|0_1\rangle\}$:
\begin{equation*}
|\Psi(t)\rangle=\sum_{n_1}\sum_{n_2>n_1}\Psi(n_1,n_2,t)|n_1,n_2\rangle~.
\end{equation*}
The transition currents are thus
\begin{align*}
&J_{(k_1,k_2)\rightarrow(k_1-1,k_2)}=-\frac{1}{\delta}\mathfrak{Re}[\Psi^*(k_1-1,k_2)\Psi(k_1,k_2)]&\\
&J_{(k_1,k_2)\rightarrow(k_1+1,k_2)}=\frac{1}{\delta}\mathfrak{Re}[\Psi^*(k_1+1,k_2)\Psi(k_1,k_2)]&\\
&J_{(k_1,k_2)\rightarrow(k_1,k_2-1)}=-\frac{1}{\delta}\mathfrak{Re}[\Psi^*(k_1,k_2-1)\Psi(k_1,k_2)]&\\
&J_{(k_1,k_2)\rightarrow(k_1,k_2+1)}=\frac{1}{\delta}\mathfrak{Re}[\Psi^*(k_1,k_2+1)\Psi(k_1,k_2)]&~,
\end{align*}
whether $k_2=k_1+1$ or not, since $\Psi(n_1,n_1)=0$. Keeping only leading terms, we get
\begin{align*}
&J_{(k_1,k_2)\rightarrow(k_1-1,k_2)}=-\frac{1}{\delta}\mathfrak{Re}[\Psi^*(k_1+1,k_2)\Psi(k_1,k_2)]=-J_{(k_1,k_2)\rightarrow(k_1+1,k_2)}&\\
&J_{(k_1,k_2)\rightarrow(k_1,k_2-1)}=-\frac{1}{\delta}\mathfrak{Re}[\Psi^*(k_1,k_2+1)\Psi(k_1,k_2)]=-J_{(k_1,k_2)\rightarrow(k_1,k_2+1)}&~.
\end{align*}
Thus there are only two transitions allowed, since the four
transitions can be arranged by pairs whose currents have opposite
signs. If $k_1$ and $k_2$ are even and correspond to $x_1$ and
$x_2$ in the continuum limit, the two transition currents are the
two discrete components of the current
\begin{equation*}
\vec{J}_{ee}(x_1,x_2,t)=\begin{pmatrix}
\mathfrak{Re}[\langle\Psi(t)|\psi^\dagger_2(x_1)\psi^\dagger_1(x_2)|0_1\rangle\langle 0_1|\psi_1(x_2)\psi_1(x_1)|\Psi(t)\rangle]\\
\mathfrak{Re}[\langle\Psi(t)|\psi^\dagger_1(x_1)\psi^\dagger_2(x_2)|0_1\rangle\langle
0_1|\psi_1(x_2)\psi_1(x_1)|\Psi(t)\rangle]
\end{pmatrix}~.
\end{equation*}Again, it is thus deterministic, since the cases odd-odd and even-odd are similar.
There are four couples that correspond to the same $(x_1,x_2)$
($(k_1,k_2)$, that we have already considered, $(k_1+1,k_2)$,
($(k_1,k_2+1)$, and ($(k_1+1,k_2+1)$). Summing their currents, we
get
\begin{equation*}
\vec{J}(x_1,x_2,t)=\begin{pmatrix}
\displaystyle\sum_{a}\langle\Psi(t)|\psi^\dagger(x_1)\psi^\dagger_a(x_2)|0_1\rangle\alpha\langle 0_1|\psi_a(x_2)\psi(x_1)|\Psi(t)\rangle\\
\displaystyle\sum_{a}\langle\Psi(t)|\psi^\dagger_a(x_1)\psi^\dagger(x_2)|0_1\rangle\alpha\langle
0_1|\psi(x_2)\psi_a(x_1)|\Psi(t)\rangle
\end{pmatrix}~.
\end{equation*}
There is also another way to prove that the model is deterministic
in the continuum limit. We take the continuum limit and then make
a rotation in the configuration space of dimension $2$, in order
to align the axis $X_1$ along a preferred direction, for example
that of $\vec{J}_{ee}(x_1,x_2,t)$, if we consider that case. Then
it can be shown that there are two transitions allowed, namely the
two transitions along $\vec{J}_{ee}(x_1,x_2,t)$. The two currents
have equal magnitude but opposite signs, so only one transition
remains. Then the particle moves in a deterministic way and the
scheme can be applied again for time $t+dt$, and so on.
\subsection{Generalization}
Suppose we have $\omega$ quanta
\begin{equation*}
\sum_{l}\phi^\dagger(l)\phi(l)|\Psi(t)\rangle=\omega|\Psi(t)\rangle~.
\end{equation*}
and an initial beable
$n_l(t_0)=\delta_{lk_1}+\delta_{lk_2}+\ldots+\delta_{lk_\omega}$.
In the continuum limit, if $x_j$ is the coordinate corresponding
to $k_j$, we expect to get the following current for the j-th
coordinate:
\begin{equation*}
J_j(t)=\sum_{s}\sum_{s_1}\ldots\sum_{s_\omega}\Psi^*_{s_1\ldots
s_j\ldots s_\omega}(t,x_1,\ldots,x_\omega)\alpha_{s_j s}
\Psi_{s_1\ldots s\ldots s_\omega}(t,x_1,\ldots,x_\omega)~,
\end{equation*}
where
\begin{equation*}
\Psi_{s_1\ldots s_\omega}(t,x_1,\ldots,x_\omega)=\langle
0_1|\psi_{s_\omega}(x_\omega)\ldots\psi_{s_1}(x_1)|\Psi(t)\rangle~.
\end{equation*}
It is just a refinement of the two-quanta case. In fact, there
will be $2\omega$ transitions allowed (to first neighbor sites).
Those transitions can be arranged by pairs. In each pair, the two
currents have equal magnitude but opposite signs. Thus there
remains only $\omega$ transitions (see Eq (\ref{tc})). Those
$\omega$ currents are just the discrete components of a continuum
current in a configuration space of dimension $\omega$. The
deterministic character can also be proved by making a rotation in
the configuration space of dimension $\omega$, as it was explained
for the two quanta case.
\subsection{The continuum limit right from the start}
We have a physical state $|\Psi(t)\rangle$ which evolves according
to the Schr\"odinger equation and we know that there is an integer
$\omega$ such that
\begin{equation*}
\int
d^3\vec{x}\psi^\dagger(\vec{x})\psi(\vec{x})|\Psi(t)\rangle=\omega|\Psi(t)\rangle~.
\end{equation*}
Thus $|\Psi(t)\rangle$ can be decomposed along eigenstates of the
fermion number density with fermion number equal to $\omega$;
those eigenstates are
\begin{equation*}
\left\{\psi^\dagger_{s_1}(\vec{x_1})\ldots\psi^\dagger_{s_\omega}(\vec{x_\omega})
|0_1\rangle~~\vec{x}_1,\ldots,\vec{x}_\omega\in R^3~~
s_1,\ldots,s_\omega\in \{1,2,3,4\}\right\}~.
\end{equation*}
Thus
\begin{equation*}
|\Psi(t)\rangle=\frac{1}{\omega!}\sum_{s_1=1}^{s_1=4}\ldots\sum_{s_\omega=1}^{s_\omega=4}\int
d^3\vec{x}_1\ldots d^3\vec{x}_\omega\Psi_{s_1\ldots
s_\omega}(t,\vec{x}_1,\ldots,\vec{x}_\omega)\psi^\dagger_{s_1}(\vec{x}_1)\ldots\psi^\dagger_{s_\omega}(\vec{x}_\omega)
|0_1\rangle~,
\end{equation*}
where the wave function is antisymmetric.
The universe at time $t$ is described by a point in a
configuration space of dimension $3\omega$ and by a pilot-state
$|\Psi(t)\rangle$. In the standard interpretation, the probability
density to observe the universe in a configuration
$(\vec{x}_1,\ldots,\vec{x}_\omega)$ is
\begin{eqnarray*}
\rho_t(\vec{x}_1,\ldots,\vec{x}_\omega)=
\sum_{s_1=1}^{s_1=4}\ldots\sum_{s_\omega=1}^{s_\omega=4}|\langle\Psi(t)|\psi^\dagger_{s_1}(\vec{x_1})\ldots\psi^\dagger_{s_\omega}
(\vec{x_\omega})|0_1\rangle|^2~,
\end{eqnarray*}
and we have
\begin{equation*}
\int d^3{\vec{x}_1}\ldots
d^3{\vec{x}_\omega}\rho_t(\vec{x}_1,\ldots,\vec{x}_\omega)=1~.
\end{equation*}
The time derivative gives a probability density current in the
configuration space:
\begin{align*}
&\frac{d}{dt}\int d^3{\vec{x}_1}\ldots
d^3{\vec{x}_\omega}\rho_t(\vec{x}_1,\ldots,\vec{x}_\omega)=
\sum_{s_1=1}^{s_1=4}\ldots\sum_{s_\omega=1}^{s_\omega=4}\int
d^3\vec{x}_1\ldots d^3\vec{x}_\omega&\\&\frac{d}{dt}[
\langle\Psi(t_0)|\psi^\dagger_{s_1}(\vec{x_1},t)\ldots\psi^\dagger_{s_\omega}(\vec{x}_\omega,t)|0_1\rangle
\langle
0_1|\psi_{s_\omega}(\vec{x}_\omega,t)\ldots\psi_{s_1}(\vec{x_1},t)|\Psi(t_0)\rangle]=0&~,
\end{align*}
where we have switched to the Heisenberg picture. It can be
simplified, knowing that
\begin{equation*}
i\frac{d\psi(t,x)}{dt}=-i\vec{\alpha}\cdot\nabla\psi(t,x)+m\beta\psi(t,x)~.
\end{equation*}
Terms containing $\beta$ cannot contribute. Thus we obtain the
following current for the j-th coordinate:
\begin{eqnarray*}
\vec{J}_j(\vec{x}_1,\ldots,\vec{x}_\omega,t)=\sum_{s=1}^{s=4}\sum_{s_1=1}^{s_1=4}\ldots\sum_{s_\omega=1}^{s_\omega=4}
\langle\Psi(t_0)|\psi^\dagger_{s_1}(\vec{x_1},t)\ldots\psi^\dagger_{s_j}(\vec{x_j},t)
\ldots\psi^\dagger_{s_\omega}(\vec{x_\omega},t)|0_1\rangle\\
\alpha_{{s_j}s}\langle
0_1|\psi_{s_\omega}(\vec{x_\omega},t)\ldots\psi_{s}(\vec{x_j},t)\ldots\psi_{s_1}(\vec{x_1},t)|\Psi(t_0)\rangle~.
\end{eqnarray*}
Define $\vec{A}=(\vec{a}_1,\ldots,\vec{a}_\omega)$, $\forall
\vec{a}\in R^3$. If $\vec{X}(t)$ is the position of the universe
in the configuration space, at time $t$, its velocity is thus
\begin{equation*}
\frac{\vec{J}(\vec{X},t)}{\rho_t(\vec{X})}\Bigg|_{\vec{X}=\vec{X}(t)}~.
\end{equation*}
\section{QUANTUM NON-LOCALITY}
The Bell model is non-local, but this is a necessary property of
any realistic interpretation of quantum field theory, following
the EPR paradox, Bell's inequality and experiments. We just want
to show it explicitly. Consider the case of two negative charges
moving in the positronic sea. Then there is an interaction among
these two charges, by the Pauli Principle; the wave function
$\Psi_{{s_1}{s_2}}(t,x_1,x_2)$ is antisymmetric. The least
entangled state, satisfying the antisymmetry requirement, is
\begin{equation}\label{las}
\Psi_{{s_1}{s_2}}(t,x_1,x_2)=\chi_{s_1}(t,x_1)\Phi_{s_2}(t,x_2)-
                                         \Phi_{s_1}(t,x_1)\chi_{s_2}(t,x_2)~.
\end{equation}
The currents are
\begin{align*}
&J_1(t,x_1,x_2)=\sum_{s_2}[\Psi^*_{1{s_2}}(t,x_1,x_2)\Psi_{2{s_2}}(t,x_1,x_2)+\Psi^*_{2{s_2}}(t,x_1,x_2)\Psi_{1{s_2}}(t,x_1,x_2)]&~,\\
&J_2(t,x_1,x_2)=\sum_{s_1}[\Psi^*_{{s_1}1}(t,x_1,x_2)\Psi_{{s_1}2}(t,x_1,x_2)+\Psi^*_{{s_1}2}
(t,x_1,x_2)\Psi_{{s_1}1}(t,x_1,x_2)]&~.
\end{align*}
Substituting $\Psi_{{s_1}{s_2}}(t,x_1,x_2)$ by the right-hand part of Eq. (\ref{las}), we get, for $J_1$:
\begin{align*}
J_1(t,x_1,x_2)=&(\chi^*_1\chi_2+\chi^*_2\chi_1)(t,x_1)(|\Phi_1|^2+|\Phi_2|^2)(t,x_2)&\\
+&(\Phi^*_1\Phi_2+\Phi^*_2\Phi_1)(t,x_1)(|\chi_1|^2+|\chi_2|^2)(t,x_2)&\\
-&(\chi^*_1\Phi_2+\chi^*_2\Phi_1)(t,x_1)(\Phi^*_1\chi_1+\Phi^*_2\chi_2)(t,x_2)&\\
-&(\Phi^*_1\chi_2+\Phi^*_2\chi_1)(t,x_1)(\chi^*_1\Phi_1+\chi^*_2\Phi_2)(t,x_2)&~.
\end{align*}
This general form is inconsistent with the existence of two real currents $J^A_1$ and $J^B_1$ such that
\begin{equation*}
J_1(t,x_1,x_2)=J^A_1(t,x_1)J^B_1(t,x_2)~,
\end{equation*}
thus the model is clearly non-local: the velocity of one of the
charges, at time $t$, depends on its position, on the pilot-state,
as well as on the position of the other negative charge, at the
same time.
\section{QUANTUM FIELD THEORY IN A FIXED SECTOR OF THE FOCK SPACE}
It is interesting to note that quantum field theory calculations
can be done in the same way as they are made in non-relativistic
quantum mechanics, at least if we only consider fermions. Let's
consider the following model, fermions interacting through a
quartic term:
\begin{equation*}
  H=\int d^3\vec{x}\left(\psi^\dagger(\vec{x})[-i\vec{\alpha}\cdot\nabla+m\beta]
  \psi(\vec{x})+g(\psi^\dagger(\vec{x})\beta\psi(\vec{x}))^2
\right)~.
\end{equation*}
Assume that there are two negative charges in the positronic sea:
\begin{equation*}
\int  d^3\vec{x}\psi^\dagger(\vec{x})\psi(\vec{x})|\Psi(t)\rangle=2|\Psi(t)\rangle~,
\end{equation*}
with
\begin{equation*}
i\frac{d|\Psi(t)\rangle}{dt}=H|\Psi(t)\rangle~.
\end{equation*}
Then the pilot-state can be decomposed along eigenstates of the
fermion number density, with fermion number equal to two:
\begin{equation*}
|\Psi(t)\rangle=\frac{1}{2!}\sum_{s_1=1}^{s_1=4}\sum_{s_2=1}^{s_2=4}\int
d^3\vec{x}_1 d^3\vec{x}_2
\Psi_{{s_1}{s_2}}(t,\vec{x}_1,\vec{x}_2)\psi^\dagger_{s_1}(\vec{x}_1)\psi^\dagger_{s_2}(\vec{x}_2)|0_1\rangle~.
\end{equation*}
Substituting the pilot-state by the right-hand part of the
previous equation in the Schr\"odinger equation, and projecting
onto a state
$\psi^\dagger_{s_1}(\vec{x}_1)\psi^\dagger_{s_2}(\vec{x}_2)|0_1\rangle$,
we get
\begin{align*}
i\frac{d}{dt}\Psi_{{s_1}{s_2}}(t,\vec{x}_1,\vec{x}_2)=
&(\beta\Psi(\vec{x}_1,\vec{x}_2))_{s_1 s_2}-(\beta\Psi(\vec{x}_2,\vec{x}_1))_{s_2 s_1}&\\
-&i(\alpha\cdot\nabla_{\vec{x}_1}\Psi_t(\vec{x}_1,\vec{x}_2))_{s_1 s_2}
+i(\alpha\cdot\nabla_{\vec{x}_2}\Psi_t(\vec{x}_2,\vec{x}_1))_{s_2 s_1}&\\
+&(\beta\Psi(t,\vec{x}_1,\vec{x}_2)\beta^{T})_{s_1 s_2}\delta(\vec{x}_1-\vec{x}_2)
-(\beta\Psi(t,\vec{x}_2,\vec{x}_1)\beta^{T})_{s_2 s_1}\delta(\vec{x}_1-\vec{x}_2)&~,
\end{align*}
where we have dropped an infinite constant.
\section{CONCLUSION}
We have obtained the continuum limit of the Bell model, for
fermions living in a one-dimensional space, using a staggered
lattice and we have also shown that we could build the continuum
Bell model directly. Physically, it is a theory of negative
charges moving in a positronic sea. There is an underlying
assumption about the state of the universe, namely that it is an
eigenstate of the fermion number (which is always true), with a
finite eigenvalue. That follows naturally from the Bell model
itself. Can one build a similar interpretation for the
Klein-Gordon quantum field theory? It seems that the answer is no,
for it is impossible to define a state annihilated by a charge
creator in the Klein-Gordon quantum field theory. Another point
worth mentioning is that the construction of the Bell model has
nothing to do with the equation of motion being linear. We could
use a Van der Waerden field and obtain the same results. Only the
Pauli exclusion principle is at work.
\begin{acknowledgments}
The author would like to thank Jean Bricmont and Thomas Durt, for
taking the time to discuss the main ideas with him.
\end{acknowledgments}
\newpage
\appendix
\section{\label{sec:appa}COMMUTATOR $[N,\psi^\dagger(\vec{x})\psi(\vec{x})]$}
We want to show that
\begin{equation}
[N,\psi^\dagger(\vec{x})\psi(\vec{x})]\neq 0~,
\end{equation}
with
\begin{equation}
N=\sum_{r}\int{d^3\vec{k}}[c^\dagger_r(\vec{k})c_r(\vec{k})+d^\dagger_r(\vec{k})d_r(\vec{k})]~.
\end{equation}
We use the following relation ($F$ stands for fermion):
\begin{align}
[F_1F_2,F_3F_4]=&F_1[F_2,F_3F_4]+[F_1,F_3F_4]F_2&\\
=&F_1\{F_2,F_3\}F_4-F_1F_3\{F_2,F_4\}+\{F_1,F_3\}F_4F_2-F_3\{F_1,F_4\}F_2&~.
\end{align}
Let's recall the expressions of the spinor fields:
\begin{align}
&\psi(\vec{x})=\sqrt{\frac{1}{(2\pi)^3}}\sum_{s}\int
{d^3\vec{p}}\sqrt{\frac{m}{E_{\vec{p}}}}[
u_s(\vec{p})e^{i\vec{p}\cdot\vec{x}}c_s(\vec{p})+
v_s(\vec{p})e^{-i\vec{p}\cdot\vec{x}}d^\dagger_s(\vec{p})]&\\
&\psi^\dagger(\vec{x})=\sqrt{\frac{1}{(2\pi)^3}}\sum_{s}\int
{d^3\vec{p}}\sqrt{\frac{m}{E_{\vec{p}}}}[
u^\dagger_s(\vec{p})e^{-i\vec{p}\cdot\vec{x}}c^\dagger_s(\vec{p})+
v^\dagger_s(\vec{p})e^{i\vec{p}\cdot\vec{x}}d_s(\vec{p})]&~.
\end{align}
By using the anti-commutation relations
\begin{align}
&\{c_s(\vec{k}),c^\dagger_r(\vec{p})\}=\delta_{sr}\delta^3(\vec{k}-\vec{p})&
&\{d_s(\vec{k}),d^\dagger_r(\vec{p})\}=\delta_{sr}\delta^3(\vec{k}-\vec{p})&~,
\end{align}
and all other anti-commutators vanishing, we find that
\begin{align}
&\{\psi^\dagger_a(\vec{x}),c_r(\vec{k})\}=\sqrt{\frac{1}{(2\pi)^3}}
\sqrt{\frac{m}{E_{\vec{k}}}}u^\dagger_{a r}(\vec{k})e^{-i\vec{k}\cdot\vec{x}}&
&\{\psi_a(\vec{x}),c_r(\vec{k})\}=0&\\
&\{\psi_a(\vec{x}),c^\dagger_r(\vec{k})\}=\sqrt{\frac{1}{(2\pi)^3}}
\sqrt{\frac{m}{E_{\vec{k}}}}u_{a r}(\vec{k})e^{i\vec{k}\cdot\vec{x}}&
&\{\psi^\dagger_a(\vec{x}),c^\dagger_r(\vec{k})\}=0&\\
&\{\psi_a(\vec{x}),d_r(\vec{k})\}=\sqrt{\frac{1}{(2\pi)^3}}
\sqrt{\frac{m}{E_{\vec{k}}}}v_{a r}(\vec{k})e^{-i\vec{k}\cdot\vec{x}}&
&\{\psi^\dagger_a(\vec{x}),d_r(\vec{k})\}=0&\\
&\{\psi^\dagger_a(\vec{x}),d^\dagger_r(\vec{k})\}=\sqrt{\frac{1}{(2\pi)^3}}
\sqrt{\frac{m}{E_{\vec{k}}}}v^\dagger_{a
r}(\vec{k})e^{i\vec{k}\cdot\vec{x}}&
&\{\psi_a(\vec{x}),d^\dagger_r(\vec{k})\}=0&~,
\end{align}
so that
\begin{align}
&[\psi^\dagger_a(\vec{x})\psi_a(\vec{x}),\sum_{r}\int{d^3\vec{k}}c^\dagger_r(\vec{k})c_r(\vec{k})]=&\\
&\sum_{r}\int{d^3\vec{k}}\bigl(\psi^\dagger_a(\vec{x})\{\psi_a(\vec{x}),c^\dagger_r(\vec{k})\}
c_r(\vec{k})
-c^\dagger_r(\vec{k})\{\psi^\dagger_a(\vec{x}),c_r(\vec{k})\}
\psi_a(\vec{x})\bigr)=&\\
&\frac{m^2}{(2\pi)^3}\sum_{s,r}\int
\frac{{d^3\vec{p}}{d^3\vec{k}}}{\sqrt{E_{\vec{p}}E_{\vec{k}}}}[
u^\dagger_s(\vec{p})u_r(\vec{k})e^{-i(\vec{p}-\vec{k})\cdot\vec{x}}c^\dagger_s(\vec{p})c_r(\vec{k})+
v^\dagger_s(\vec{p})u_r(\vec{k})e^{i(\vec{p}+\vec{k})\cdot\vec{x}}d_s(\vec{p})c_r(\vec{k})]-
&\\
&\frac{m^2}{(2\pi)^3}\sum_{s,r}\int
\frac{{d^3\vec{p}}{d^3\vec{k}}}{\sqrt{E_{\vec{p}}E_{\vec{k}}}}[
u^\dagger_r(\vec{k})u_s(\vec{p})e^{i(\vec{p}-\vec{k})\cdot\vec{x}}
c^\dagger_r(\vec{k})c_s(\vec{p})+
u^\dagger_r(\vec{k})v_s(\vec{p})e^{-i(\vec{p}+\vec{k})\cdot\vec{x}}
c^\dagger_r(\vec{k})d^\dagger_s(\vec{p})] &~.
\end{align}
Since $r$, $s$, $\vec{p}$ and $\vec{k}$ are dummy variables, we find that
\begin{align}
&[\psi^\dagger_a(\vec{x})\psi_a(\vec{x}),\sum_{r}\int{d^3\vec{k}}c^\dagger_r(\vec{k})c_r(\vec{k})]=&\\
&\frac{m^2}{(2\pi)^3}\sum_{s,r}\int
\frac{{d^3\vec{p}}{d^3\vec{k}}}{\sqrt{E_{\vec{p}}E_{\vec{k}}}}[
v^\dagger_s(\vec{p})u_r(\vec{k})e^{i(\vec{p}+\vec{k})\cdot\vec{x}}d_s(\vec{p})c_r(\vec{k})]-
&\\
&\frac{m^2}{(2\pi)^3}\sum_{s,r}\int
\frac{{d^3\vec{p}}{d^3\vec{k}}}{\sqrt{E_{\vec{p}}E_{\vec{k}}}}[
u^\dagger_r(\vec{k})u_s(\vec{p})e^{i(\vec{p}-\vec{k})\cdot\vec{x}}
c^\dagger_r(\vec{k})c_s(\vec{p})] &~.
\end{align}
In the same way, we obtain
\begin{align}
&[\psi^\dagger_a(\vec{x})\psi_a(\vec{x}),\sum_{r}\int{d^3\vec{k}}d^\dagger_r(\vec{k})d_r(\vec{k})]=&\\
&\sum_{r}\int{d^3\vec{k}}\bigl(-\psi^\dagger_a(\vec{x})d^\dagger_r(\vec{k})
\{\psi_a(\vec{x}),d_r(\vec{k})\}+\{\psi^\dagger_a(\vec{x}),d^\dagger_r(\vec{k})\}d_r(\vec{k})
\psi_a(\vec{x})\bigr)=-&\\
&\frac{m^2}{(2\pi)^3}\sum_{s,r}\int
\frac{{d^3\vec{p}}{d^3\vec{k}}}{\sqrt{E_{\vec{p}}E_{\vec{k}}}}[
u^\dagger_s(\vec{p})v_r(\vec{k})e^{-i(\vec{p}+\vec{k})\cdot\vec{x}}c^\dagger_s(\vec{p})
d^\dagger_r(\vec{k})+
v^\dagger_s(\vec{p})v_r(\vec{k})e^{i(\vec{p}-\vec{k})\cdot\vec{x}}d_s(\vec{p})d^\dagger_r(\vec{k})]+
&\\
&\frac{m^2}{(2\pi)^3}\sum_{s,r}\int
\frac{{d^3\vec{p}}{d^3\vec{k}}}{\sqrt{E_{\vec{p}}E_{\vec{k}}}}[
v^\dagger_r(\vec{k})u_s(\vec{p})e^{i(\vec{p}+\vec{k})\cdot\vec{x}}
d_r(\vec{k})c_s(\vec{p})+
v^\dagger_r(\vec{k})v_s(\vec{p})e^{-i(\vec{p}-\vec{k})\cdot\vec{x}}
d_r(\vec{k})d^\dagger_s(\vec{p})] &~.
\end{align}
This can be simplified to
\begin{align}
&[\psi^\dagger_a(\vec{x})\psi_a(\vec{x}),\sum_{r}\int{d^3\vec{k}}d^\dagger_r(\vec{k})d_r(\vec{k})]=&\\
&\frac{m^2}{(2\pi)^3}\sum_{s,r}\int
\frac{{d^3\vec{p}}{d^3\vec{k}}}{\sqrt{E_{\vec{p}}E_{\vec{k}}}}[
u^\dagger_s(\vec{p})v_r(\vec{k})e^{-i(\vec{p}+\vec{k})\cdot\vec{x}}c^\dagger_s(\vec{p})
d^\dagger_r(\vec{k})]+
&\\
&\frac{m^2}{(2\pi)^3}\sum_{s,r}\int
\frac{{d^3\vec{p}}{d^3\vec{k}}}{\sqrt{E_{\vec{p}}E_{\vec{k}}}}[
v^\dagger_r(\vec{k})u_s(\vec{p})e^{i(\vec{p}+\vec{k})\cdot\vec{x}}
d_r(\vec{k})c_s(\vec{p})] &~.
\end{align}
Putting the two results together, we get
\begin{align}
&[\psi^\dagger_a(\vec{x})\psi_a(\vec{x}),N]=&\\
&\frac{2m^2}{(2\pi)^3}\sum_{s,r}\int
\frac{{d^3\vec{p}}{d^3\vec{k}}}{\sqrt{E_{\vec{p}}E_{\vec{k}}}}[
u^\dagger_s(\vec{p})v_r(\vec{k})e^{-i(\vec{p}+\vec{k})\cdot\vec{x}}c^\dagger_s(\vec{p})
d^\dagger_r(\vec{k})]+
&\\
&\frac{2m^2}{(2\pi)^3}\sum_{s,r}\int
\frac{{d^3\vec{p}}{d^3\vec{k}}}{\sqrt{E_{\vec{p}}E_{\vec{k}}}}[
v^\dagger_r(\vec{k})u_s(\vec{p})e^{i(\vec{p}+\vec{k})\cdot\vec{x}}
d_r(\vec{k})c_s(\vec{p})]
&~,
\end{align}
which is not equal to zero, even if we think about fields as distributions. If we start from the state
 $d^\dagger_s(p_0)c^\dagger_s(p_0)|0\rangle$, it is clear that there are well-behaved functions $f$ such that
\begin{equation}
\langle 0|\int
d^3\vec{x}f(\vec{x})[\psi^\dagger(\vec{x})\psi(\vec{x}),N]|d^\dagger_s(p_0)c^\dagger_s(p_0)|0\rangle\neq
0~.
\end{equation}
\newpage %Just because of unusual number of tables stacked at end
\bibliography{bqft}% Produces the bibliography via BibTeX.
\end{document}